\documentstyle[eqsecnum,prd,aps,preprint,tighten]{revtex}
\newcommand{\beq}{\begin{equation}}
\newcommand{\eeq}{\end{equation}}
\newcommand{\bea}{\begin{eqnarray}}
\newcommand{\eea}{\end{eqnarray}}

 \def\t{\tilde}
\def\m{\mu}
\def\n{\nu}

\def\G{\Gamma}
\def\na{\nabla}
\def\muhat{\hat{\mu}}
\def\nuhat{\hat{\nu}}

\newcommand{\no}{\nonumber \\}

\def\CN{{\cal N}}

\preprint{\hfill\parbox{4cm}{UT-KOMABA 00-09}}

\begin{document}
\draft \title{Three point functions and the effective lagrangian for the chial primary fields in $D=4$ supergravity on $AdS_2 \times S^2 $}
\author{Julian Lee\footnote{jul@hep1.c.u-tokyo.ac.jp}}
\address{Institute of Physics, University of Tokyo \\                
 Komaba, Meguro-ku, Tokyo 153, Japan}

\date{\today}
\maketitle 
\begin{abstract}
 We consider the $D=4$, $N=8$ supergravity on $AdS_2 \times S^2$ space. We obtain the truncated Lagrangian for the bosonic chiral primary fields, and compute the tree level three-point correlation functions. 
\end{abstract}

\setcounter{footnote}{0}
\narrowtext 

\newpage
\section{Introduction}
 Among all known examples of the $AdS$/CFT correspondence 
\cite{malda}, \cite{gkp}, \cite{witt},\cite{rev}, 
the least understood is the $AdS_2$/CFT$_1$ case. 
The $D=1$ conformal field theory (CFT), or conformal quantum mechanics (CQM), 
has not been formulated and therefore no quantitative comparison
between the two sides of the duality has been made.  However, there have been conjectures\cite{gibtown},\cite{town},\cite{jy} that the dual $CFT$
is given by the $n$-particle, ${\cal N} = 4$ superconformal Calogero 
model, which has yet to be constructed for arbitrary $n$\cite{AkuKud}.  By going to the second-quantized formulation, this becomes a $1+1$ dimensional field theory. (See also ref.\cite{strom}, \cite{frag}, \cite{nak}, \cite{cad}.)

As a first step toward investigating this conjecture we consider the 
supergravity  side of the duality.  The low energy limit of type II(A or B) string
theory on $T^6$ is the ${\cal N}=8$ supergravity theory of 
Cremmer and Julia \cite{CremJul}.  Off-shell this theory has
an $SO(8)$ symmetry while on-shell the symmetry is enhanced to an
$E_{7(7)}$ duality.  We consider the fluctuations of the fields around a classical configuration called  Bertotti-Robinson solution, which gives the $AdS_2 \times S^2$ spacetime. This solution is the near-horizon limit of four D-branes intersecting over a string. The equations for the fluctuations of the fields were considered up to linear order in ref.\cite{corley},\cite{LL} to obtain the physical spectrum of the theory. \footnote{ See also ref.\cite{spra},\cite{y2} for the spectrum of the minimal $D=4, \CN=2$ supergravity on $AdS_2 \times S^2$ and the $D=10$, IIA supergravity on `quasi' $AdS_2 \times S^8$, respectively.} 

 In this paper, we compute the cubic interactions among the bosonic chiral primary fields. Such computations were done for $D=10$ IIB supergravity on $AdS_5 \times S^5$ and $D=6$ supergravity on $AdS_3 \times S^3$ in ref.\cite{leeseiberg} and \cite{mihail}, respectively.  The computation 
involves the standard elements of previous works, namely nonlinear field 
redefinitions, which then lead to three point functions of chiral 
operators. The results  have some 
similarities in form with the examples  in the higher dimensional $AdS$ spaces. It would be interesting to see whether one can make computations in a supersymmteric Calogero model, to be compared with these results.

\section{Supergravity in $D=4$}
In this section, we review  following \cite{corley}, \cite{LL} the equations of motion of $D=4$, $\CN=8$ supergravity. We mainly follow the four dimensional formalism and notation of ref.~\cite{corley}.  We are interested in the equation of motion for the metric 
$g_{\hat \m \hat \n}$, vectors $B^{AB}_{\hat \m}$, and scalars $W_{ABCD}$, where the capital roman letters are $SU(8)$ indices.  The scalars are packaged in a $E_{7(7)}$ matrix ${\cal V}$  parametrizing  the coset ${E_{7(7)} \over {SU(8)}}$\cite{CremJul},
\beq
\partial_{\muhat}{\cal V}{\cal V}^{-1} = \pmatrix{
Q_{\muhat [A}^{\;\;\;\;\;[C} \delta_{B]}^{\;\;\;D]}  & 
P_{\muhat ABCD} \cr
{\bar P}_{\muhat}^{\;\;ABCD}  &
{\bar Q}_{\muhat \;\;\;\;\;[C}^{\;\;\;[A} 
\delta^{B]}_{\;\;\;D]}.}
\label{Vconnection}
\eeq 
 $P$'s parametrize the coset manifold and can be expressed in terms of the scalar fields $W$, and $Q$'s are the $SU(8)$ gauge fields, but for our purposes it is enough to note that the $SU(8)$ gauge symmetry can be fixed to the so called symmetric gauge where 
  ${\cal V} = \exp(X)$,
\beq
X = \pmatrix{
 0 &
 W_{ABCD} \cr
 \bar W^{ABCD}   &
 0} ,
\label{gauge}
\eeq
and $W_{ABCD}$ is complex, completely antisymmetric in $A,B,C,D$, and satisfies
the constraint
\begin{eqnarray}
\bar{W}^{ABCD} = \frac{1}{24} \epsilon^{ABCDEFGH} W_{EFGH}.
\label{Wconstraint}
\end{eqnarray}
   We will use the following notation for indices: 
$\hat \m, \hat \nu = 0,1,2,3$ are $D=4$ coordinates, $\lambda, \mu, \nu\cdots=0,1$ are $AdS_2$ coordinates 
and $\alpha,\beta,\gamma \cdots =2,3$ are $S^2$ coordinates.

  The bosonic part of the $D=4$ supergravity action is:
\begin{equation}
{\cal L}  =  \biggl(\frac{1}{4} e R(\omega,e)   
+ \frac{1}{8} e
F_{\muhat \nuhat}^{MN}(B) \tilde H_{MN}^{\muhat \nuhat}(B,{\cal V})
 -  \frac{1}{24} e P_{\muhat ABCD} \bar P^{\muhat ABCD}  \biggr)
\label{Lfull}
\end{equation}
where
\beq
F_{\muhat \nuhat}^{AB}  =  2 \partial_{[\muhat} B^{AB}_{\nuhat]}  
\eeq
   $\tilde{H}^{\muhat \nuhat}_{MN}$ are defined in the Appendix 1. For our purpose, we only need  the leading expansion of the $\t H$ and $P$ in $W^{ABCD}$:
\begin{eqnarray}
G^{MN}_{\muhat \nuhat} \tilde{H}^{(B)}_{\muhat \nuhat MN} &=& 
-G^{MN}_{\muhat \nuhat} (1+W+\bar{W}+W^2+\bar{W}^2 \no 
&&-{1 \over 3} W \bar{W} W -{1 \over 3} \bar{W} W \bar{W} +  W^3 + \bar {W}^3 )_{MNPQ}  G^{\muhat \nuhat PQ} \no 
&&+ i G^{MN}_{\muhat \nuhat}(W-\bar{W}+W^2-\bar{W}^2+ O(W^3) )_{MNPQ} 
\tilde{G}^{\muhat \nuhat PQ} + O(W^4) , \no
P_{\muhat ABCD}  &=& \na_{\hat \mu} W_{ABCD} + O(W^3). 
\label{Hbosonic}
\end{eqnarray} 
 
\section{equations of motion for the bosonic chiral primary fields}
 We consider the classical configuration called Bertotti-Robinson solution\cite{berob},
\bea
ds^2 &=& 
{1\over z^2} (-dx_0^2 + dz^2) + d\Omega_2^2,\no
R_{\mu\lambda\nu\sigma} &=& -(g_{\mu\nu}g_{\lambda\sigma} - g_{\mu\sigma}g_{\lambda\nu}) \no
R_{\alpha \gamma\beta\delta} &=& (g_{\alpha\beta}g_{\gamma\delta} - g_{\alpha\delta}g_{\gamma\beta}) \no 
\bar F^{12}_{\alpha \beta} &=& \epsilon_{\alpha \beta},\no
\bar F^{AB}_{\alpha \beta} &=& 0 (A \neq 1,2),\no
W_{ABCD} &=& 0
\eea
which breaks the $SU(8)$ internal symmetry into $SU(6) \times SU(2)$. The 
bulk fields of interest are the fluctuations about this background, 
\bea
  g_{\muhat \nuhat}&=&\bar{g}_{\muhat \nuhat}+h_{\muhat \nuhat}, \no 
   F^{AB}_{\muhat \nuhat}&=&\bar{F}^{AB_{\muhat \nuhat}}+2 \na_{[\hat \mu} b^{AB}_{\hat \nu ]}
\eea 
and $W_{ABCD}$. One can consider the classical equations for these fluctuations to linear order and organize the spectrum into the multiplets of $SU(6) \times SU(2)$\cite{corley},\cite{LL}. In this paper we expand the equations of motion up to the  second order in these fluctuations to obtain the interaction terms.

 To find the chiral primary fields on $AdS_2$ we expand the fields in spherical harmonics on $S^2$.  The expansions are quite simple in this case
as all harmonic functions on the 2-sphere can be expressed in terms
of just the scalar spherical harmonics $Y_{lm}$. $l$ is the quantum number labeling the Casimir of the representation, 
\beq
 \nabla_{\alpha} \nabla^{\alpha} Y_{lm}= -l(l+1) Y_{lm}.
\eeq
 The expansions
of the bosonic fluctuations are then given by (denoting the $l,m$
indices collectively by $I$)
\begin{mathletters}
\begin{eqnarray}
h_{\mu \nu} & = & \sum H^{I}_{\mu \nu} Y_{I} \label{hmunuexp} \\
h_{\mu \alpha} & = & \sum (B^{I}_{1 \mu} \nabla_{\alpha} Y_{I}
+ B^{I}_{2 \mu} e_{\alpha \beta} \nabla^{\beta} Y_{I})
\label{hmualphaexp} \\
h_{\alpha \beta} & = & \sum (\phi^{I}_1 \nabla_{\alpha} \nabla_{\beta}
Y_{I} + \phi^{I}_2 e_{(\alpha}^{\;\;\;\gamma} \nabla_{\beta)}
\nabla_{\gamma} Y_{I} + \phi^{I}_3 g_{\alpha \beta} Y_{I})
\label{halphabetaexp} \\
b_{\mu}^{AB} & = & \sum b_{\mu}^{(I)AB} Y_{I} \label{bmuexp} \\
b_{\alpha}^{AB} 
& = & \sum (b^{(I)AB}_{1} \nabla_{\alpha} Y_I
+ b^{(I)AB}_{2} e_{\alpha \beta} \nabla^{\beta} Y_{I}) 
\label{balphaexp} \\
W_{ABCD} & = & \sum W^I_{ABCD} Y_I.
\label{Wexp}
\end{eqnarray}
\end{mathletters}

Before substituting the expansions into the  equations of
motion we can first simplify the expansions 
by fixing  gauge symmetries by imposing\footnote{Since we project out only the physical modes, actually it is enough to impose $\phi_1=\phi_2=0$. From now on, we write $\phi_3 \to \phi, B_{2 \m} \to B_\mu, b_2 \to b$. }  
\begin{equation}
\phi_{1}^I = \phi_{2}^I = B_{1 \mu}^I  = b_{1}^{(I)AB}  =  0.  
\label{diffeofix}
\end{equation}
 At the linearized level, the bosonic fields can be decomposed into the eigenstates of the $AdS_2$ Laplacian\cite{corley},\cite{LL},
\bea
\phi^I  &=& 2 l T^I + 2 (l-1) \tilde T^I \nonumber \\
b^{(I)[12]}&=& T^I - \tilde T^I \nonumber \\
a^{(I)[12]}&\equiv&\epsilon^{\mu \nu} \na_\m b^{(I)[12]}_\n =  (l+1) S^I + l \tilde S^I \nonumber \\
B^I&\equiv& \epsilon^{\mu \nu} \na_\m B^I_\nu = -2 S^I + 2 \tilde S^I \nonumber \\
a^{(I)[MN]}&\equiv&\epsilon^{\mu \nu} \na_\m b^{(I)[MN]}_\n = (l-1)  U^{(I)[MN]} + (l+2) \tilde  U^{(I)[MN]}  \quad (M, N \neq 1,2) \nonumber \\
(W-\bar W)^{(I)12MN} &=& i \na_x^2 (U^{(I)[MN]} - \tilde  U^{(I)[MN]}) \nonumber \\
 b^{(I)[MN]}&=&  V^{(I)[MN]} + \tilde  V^{(I)[MN]} \quad (M, N \neq 1,2) \nonumber \\
(W+\bar W)^{(I)12MN} &=& -l V^{(I)[MN]} + (l+1) \tilde  V^{(I)[MN]} \label{chpr}
\eea
where these fields satisify the equations of the form:
\bea
(\nabla_x^2 - l(l-1))A^I +Q_A^I&=&0 \no \\
(\nabla_x^2 - (l+1)(l+2))\tilde A^I +\tilde Q_{\tilde A}^I=0 \label{eigen}
\eea 
Here, $A$ and $\tilde A$ stand for $T,S,U,V$ and $\tilde T, \tilde S, \tilde U, \tilde V$ respectively, and  $Q_A$ and $Q_{\tilde A}$ are the second order corrections. Also, we note that the fields $H_{\mu \nu}$ are not independent degrees of freedom and they are completely determined by the equations\cite{corley},\cite{LL},\cite{spra}
\begin{mathletters}
\begin{eqnarray}
\Bigl( \bigl(\nabla^{2}_x +2-l(l+1) \bigr) 
H^I_{\mu \nu} -2 \nabla_{(\mu}
\nabla^{\lambda} H^I_{\nu) \lambda} + 
\bigl(\nabla_{\mu} \nabla_{\nu}
&-& g_{\mu \nu} (\nabla^{2}_x +1  -  l(l+1)) \bigr) H^I \no + g_{\mu \nu}
\nabla^{\lambda} \nabla^{\rho} H^I_{\lambda \rho}
+ 2 \bigl(
\nabla_{\mu} \nabla_{\nu} - g_{\mu \nu} (\nabla^{2}_x -1-\frac{1}{2}l(l+1))
\bigr) \phi^I \Bigr) & = & 4 g_{\mu \nu} l(l+1) b^I + {\rm (higher\  order)} 
\label{Emunu} \\
\bigl( \nabla^{\nu} H^I_{\mu \nu} - \nabla_{\mu} H^I - \nabla_{\mu}
\phi^I \bigr) & = & -4 \nabla_{\mu} b^I + {\rm (higher\ order)}
\label{Emualpha2} \\
\Bigl( \bigl( \nabla^{2}_x + 4 \bigr) \phi^I + 
\bigl(\nabla^{2}_x -1-l(l+1) \bigr)
H^I - \nabla^{\mu} \nabla^{\nu} H^I_{\mu \nu} \Bigr) & = & 
4 l(l+1) b^I+ {\rm (higher\ order)} \label{Ealphabeta1} \\
H^I & = & {\rm (quadratic\ order)}. \label{Ealphabeta2}
\end{eqnarray}
\end{mathletters}
By making an ansatz, one can easily find the solution to the equations above:
\beq
H^I_{\mu \nu} = {1 \over l+1} \left(-2 l (l-1) g_{\mu \nu} T^I + 4 \na_\mu \na_\nu  T^I \right) + {\rm (higher\ order\ corrections) }\label{meq}
\eeq
Since we are interested only in the three-point function of chiral primary fields, only $Q_T, Q_S, Q_U, Q_V$ are of interest.  We also put any non-chiral primary fields appearing in $Q$'s to be zero, and substitute (\ref{meq}) for $H_{\mu \nu}$, neglecting the higher order corrections.   The detailed form of the $Q$'s are  rather complicated and not interesting at this stage. They are of the generic form:
\beq
Q_A =  \alpha \nabla^\mu \na_\nu B\na_\mu \na_\nu C + \beta \na^\mu B \na_\mu C + \gamma B C  + \cdots, 
\eeq
where $A$, $B$, $C$ are chiral primary fields. 
   We see that there are terms involving derivatives of the fields in $Q$'s. These terms can be removed by nonlinear redefinitions of the fields, which does not change the equations (\ref{eigen}) at the linear level. It is easy to see that this can be done by redefining\cite{leeseiberg},\cite{mihail} 
\beq
A \to A - {\alpha \over 2 c} (\na B) \cdot (\na C) - {1 \over 2 c} ( \alpha + (1+\Gamma) \beta) B \cdot C,
\eeq
where $\Gamma \equiv {1 \over 2}(l(l-1)-l_1(l_1-1)-l_2(l_2-1))$ with $l,l_1,l_2$ being the total angular momentum quantum numbers of $A,B,C$, respectively. 
 After these redefinition, the linear term 
\beq
(\na_x^2 - l(l-1))A
\eeq
generates additional term which removes the derivative terms, and $Q_A$ becomes:
\beq
Q_A = (\gamma +  \Gamma ( \beta + (1 + \Gamma) \alpha)) B C\cdots     
\eeq
We then get
\bea
Q_T &=& { (l(l^2-1)+l_1(l_1^2-1)+l_2(l_2^2-1)) 
         \over
      2 l (2l+1)(l-1) (l_1+1) (l_2+1)}\alpha \alpha_1 \alpha_2 (\Sigma^2-1) \tilde C(I; I_1,I_2) T^2 \no 
&& - { (l(l^2-1)-l_1(l_1^2-1)-l_2(l_2^2-1)) \tilde C(I; I_1,I_2) \over 2 l(2l+1)(l-1)} \alpha \alpha_1 \alpha_2 (\Sigma^2-1) S^2  \no 
&&+{(l_1 + l_2)(l_1-1)(l_2-1)  \over 4 l (2l+1)(l-1) } \alpha \alpha_1 \alpha_2 
       (\Sigma^2-1) \tilde C(I; I_1,I_2) \epsilon^{12ABCDEF} U_{CD} U_{EF} \no 
&&+{ (l_1 + l_2) \over 4 l(2l+1)(l-1)}\alpha \alpha_1 \alpha_2
     (\Sigma^2-1)  \tilde C(I; I_1,I_2) \epsilon^{12ABCDEF} V_{CD} V_{EF} \\
\no 
 Q_S &=&  {(l(l^2-1)+l_1(l_1^2-1)-l_2(l_2^2-1))  \over 2l(l^2-1)(2l+1)(l_2+1)}\alpha \alpha_1 \alpha_2  (\Sigma^2-1)\tilde C(I; I_1,I_2) S^{(1)} T^{(2)} \no 
&&- {(l_1 - l_2)(l_2-1) \over 4 l(l^2-1)(2l+1)} \alpha \alpha_1 \alpha_2 (\Sigma^2-1) \tilde C(I; I_1,I_2) V^{(1)} U^{(2)} \\
\no 
 Q^{AB}_U &=& {(l - l_2)  \over 2l (l-1)(2l+1)}\alpha \alpha_1 \alpha_2   (\Sigma^2-1)\tilde C(I; I_1,I_2) S^{(1)} V^{(2)AB} \no 
        &&-{l_1-1 \over 8 l (l-1)(2l+1)}  \alpha \alpha_1 \alpha_2  (\Sigma^2-1)\tilde C(I; I_1,I_2)\epsilon^{1 2 A B C D E F} U^{(1)}_{CD}V^{(2)}_{EF} \no 
&& +{ (l + l_1)(l_1-1)  \over 2 l(l-1)(2l+1) (1 + l_2)}\alpha \alpha_1 \alpha_2
       (\Sigma^2-1)\tilde C(I; I_1,I_2) U^{(1) AB}T^{(2)} \\
\no 
Q^{AB}_V &=&  {(l_2-l)(l_2-1)  \over 2 l(2l+1)}
       \alpha \alpha_1 \alpha_2  (\Sigma^2-1) \tilde C(I; I_1,I_2)  S^{(1)}U^{(2)AB }\no  
&&   -{(l_1-1)(l_2-1)  \over 16 l(2l+1) }\alpha \alpha_1 \alpha_2
     (\Sigma^2-1) \tilde C(I; I_1,I_2) \epsilon^{1 2 A B C D E F}U_{C D} U_{E F} \no 
         && +{1 \over 16 l(2l+1)} \alpha \alpha_1 \alpha_2  (\Sigma^2-1) \tilde C(I; I_1,I_2) \epsilon^{1 2 A B C D E F}V_{C D} V_{E F} \no 
         && +{(l + l_1) \over  2 l (2l +1) (1 + l_2) } \alpha \alpha_1 \alpha_2  (\Sigma^2-1) \tilde C(I; I_1,I_2)  V^{(1)AB} T^{(2)} \\ \nonumber
\eea
at the quadratic level, where $\alpha \equiv l_1 + l_2 -l$, $\alpha_1 \equiv l+l_2 -l_1$, $\alpha_2 \equiv l + l_1 - l_2$, $\Sigma \equiv l+l_1+l_2$, and
\beq
\tilde C(I; I_1, I_2) \equiv \int Y_I^* Y_{I_1} Y_{I_2}.
\eeq
We also used the abbreviation $A^{(i)} \equiv A^{I_i}$ and $A^2 \equiv A^{I_1} A^{I_2}$ for any field $A$. 
\section{The effective action and the three-point function function}
The equations  in the previous section can be derived from the truncated Lagrangian
\bea
{\cal L}  &=&  \biggl(\frac{1}{4} e R(\omega,e)   
+ \frac{1}{8} e 
F_{\muhat \nuhat}^{MN}(B) \tilde H_{MN}^{\muhat \nuhat}(B,{\cal V})
 -  \frac{1}{24} e P_{\muhat ABCD} \bar P^{\muhat ABCD}  \biggr) \no \no
&=& {(2 l+1)l (l^2-1) \over 4} T(\na_x^2 -l(l-1))T + {(2 l+1)l (l^2-1) \over 2} S(\na_x^2 -l(l-1))S \no
&&+ {l (2l+1) \over 4} V^{AB}(\na_x^2 -l(l-1))V_{AB} + {l(l-1)^2(2l+1) \over 4} U^{AB}(\na_x^2 -l(l-1))U_{AB} \no \no
&&+{(l_1(l_1^2-1)+l_2(l_2^2-1)+l_3(l_3^2-1)) 
         \over
      3 (l_1+1) (l_2+1)(l_3+1)}\alpha_1 \alpha_2 \alpha_3(\Sigma^2-1) C(I_1, I_2, I_3) T^3 \no \no
&&+{(l_1(l_1^2-1)+l_2(l_2^2-1)-l_3(l_3^2-1)) 
         \over
       (l_3+1)} \alpha_1 \alpha_2\alpha_3  (\Sigma^2-1) C(I_1, I_2,I_3) S^{(1)} S^{(2)} T^{(3)} \no \no 
&&-(l_1-l_2)(l_2-1) 
        \alpha_1 \alpha_2 \alpha_3 (\Sigma^2-1) C(I_1, I_2, I_3) V^{(1)AB} U^{(2)}_{AB} S^{(3)} \no \no
&&+{1 \over 24} \alpha_1 \alpha_2 \alpha_3 (\Sigma^2-1) C(I_1, I_2, I_3) \epsilon^{12ABCDEF} V^{(1)}_{AB} V^{(2)}_{CD} V^{(3)}_{EF} \no \no
&&+{l_1 + l_2 \over 2 (l_3+1)} \alpha_1 \alpha_2 \alpha_3 (\Sigma^2-1) C(I_1, I_2, I_3)  V^{(1)AB} V^{(2)}_{AB} T^{(3)} \no \no 
&&-{(l_1-1)(l_2-1) \over 8} \alpha_1 \alpha_2 \alpha_3 (\Sigma^2-1) C(I_1, I_2, I_3)  \epsilon^{12ABCDEF} U^{(1)}_{AB} U^{(2)}_{CD} V^{(3)}_{EF} \no \no
&&+{(l_1-1)(l_2-1)(l_1 + l_2) \over 2 (l_3 + 1)} \alpha_1 \alpha_2 \alpha_3 (\Sigma^2-1) C(I_1, I_2, I_3)  U^{(1)AB} U^{(2)}_{AB} T^{(3)}.\label{trla}
\eea
with
\beq
 C(I_1, I_2, I_3) \equiv \int Y^{I_1} Y^{I_2} Y^{I_3}.
\eeq
Here the normalizations were fixed by directly substituting the expression (\ref{chpr}) into the lagrangian (\ref{Lfull}) and evaluating the leading terms for some fields.  

To compute from (\ref{trla}) the 2- and 3-point functions of chiral primary operators of the boundary theory, we
apply the formulae derived, for instance, in ref.\cite{rmathur}. From Eq.(17) and
the correction factor in Eq.(95) of ref.\cite{rmathur}, we read off the tree-level two-point functions to be\footnote{Although the following results are the correlation functions of the chiral primary operators of the boundary theory which couple to the bulk fields $S,T, U, V$ we will still denote them by $S, T, U, V$ for the simplicity of the notations.} 
\bea
\langle T^{I_1}(x) T^{I_2}(y) \rangle &=& {(2l+1) (l^2-1)
 \over 2}
{1\over \pi^{1/2}} {\G(l+1)\over \G(l-1/2)}(2l-1)
{ \delta^{I_1 I_2} \over |x-y|^{2l}}.
 \no
\langle S^{I_1}(x) S^{I_2}(y) \rangle &=& (2l+1) (l^2-1)
 {1\over \pi^{1/2}} {\G(l+1)\over \G(l-1/2)}(2l-1)
{ \delta^{I_1I_2} \over |x-y|^{2l}}.
 \no
\langle U^{(I_1) AB}(x) U^{(I_2) CD}(y) \rangle &=& (l-1)^2(2l+1)
{1\over \pi^{1/2}} {\G(l+1)\over \G(l-1/2)}(2l-1)
{(\delta_{AC} \delta_{BD} - \delta_{AD}\delta_{BC}) \delta^{I_1I_2} \over |x-y|^{2l}}.
 \no
\langle  V^{(I_1) AB}(x) V^{(I_2) CD}(y) \rangle &=& (2l+1) 
{1\over \pi^{1/2}} {\G(l+1)\over \G(l-1/2)}(2l-1)
{(\delta_{AC} \delta_{BD} - \delta_{AD}\delta_{BC}) \delta^{I_1I_2} \over |x-y|^{2l}}.
\eea
{}From Eq.(25) of the same paper we derive that
\bea
\langle T^{(1)}(x) T^{(2)}(y) T^{(3)}(z) \rangle &=& -{2^5 \over \pi}
{\prod_i \G({\alpha_i \over 2}+1) \G({1 \over 2} \Sigma
 +{3 \over 2})(l_1(l_1^2-1)+l_2 (l_2^2-1) + l_3 (l_3^2-1)) \over 
\prod_i^3 \left( \G(l_i -1/2)(l_i +1) \right) |x-y|^{\alpha_3} |y-z|^{\alpha_1} |z-x|^{\alpha_2} }C(I_1, I_2,I_3) \no
\no
\langle S^{(1)} S^{(2)} T^{(3)}\rangle   &=&-{2^5 \over \pi}
{\prod_i \G({\alpha_i \over 2}+1 ) \G({1 \over 2} \Sigma +{3 \over 2})(l_1(l_1^2-1)+l_2 (l_2^2-1) + l_3 (l_3^2-1)) \over \prod_i^3 \left( \G(l_i -1/2) \right)(l_3+1) |x-y|^{\alpha_3} |y-z|^{\alpha_1} |z-x|^{\alpha_2} }C(I_1, I_2,I_3) \no
\no
\langle  V^{(1)AB} U^{(2)CD} S^{(3)} \rangle &=&{2^5 \over \pi}
{\prod_i \G({\alpha_i \over 2}+1) \G({1 \over 2} \Sigma +{3 \over 2})(l_1-l_2)(l_2-1) \over \prod_i^3 \left( \G(l_i -1/2) \right) |x-y|^{\alpha_3} |y-z|^{\alpha_1} |z-x|^{\alpha_2} }\no
&& \times (\delta^{AC}\delta^{BD} - \delta^{AD}\delta^{BC})C(I_1, I_2,I_3)  \no 
\no 
\langle V^{(1)}_{AB} V^{(2)}_{CD} V^{(3)}_{EF} \rangle   & = &-{2^5  \over \pi}
{\prod_i \G({\alpha_i \over 2}+1) \G({1 \over 2} \Sigma +{3 \over 2}) \over \prod_i^3 \left( \G(l_i -1/2) \right) |x-y|^{\alpha_3} |y-z|^{\alpha_1} |z-x|^{\alpha_2} }C(I_1, I_2,I_3)  \epsilon_{12ABCDEF}  \no 
\no
\langle V^{(1)}_{AB} V^{(2)}_{CD} T^{(3)} \rangle   &=& -{2^6 \over \pi}
{(l_1 + l_2)\prod_i \G({\alpha_i \over 2}+1) \G({1 \over 2} \Sigma +{3 \over 2}) (\delta_{AC} \delta_{BD} - \delta_{AD} \delta_{BC} )\over (l_3+1) \prod_i^3 \left( \G(l_i -1/2) \right) |x-y|^{\alpha_3} |y-z|^{\alpha_1} |z-x|^{\alpha_2} }C(I_1, I_2,I_3)   \no 
\no 
\langle  U^{(1)}_{AB} U^{(2)}_{CD} V^{(3)}_{EF} \rangle   &=&{2^5 \over \pi}
{(l_1-1)(l_2-1)\prod_i \G({\alpha_i \over 2}+1) \G({1 \over 2} \Sigma +{3 \over 2}) \over \left(  \prod_i^3 \G(l_i -1/2)  \right) |x-y|^{\alpha_3} |y-z|^{\alpha_1} |z-x|^{\alpha_2} }C(I_1, I_2,I_3)\epsilon_{12ABCDEF} \no \no
\langle  U^{(1)AB} U^{(2)CD} T^{(3)} \rangle    &=& -{2^6 \over \pi}
{(l_1-1)(l_2-1)(l_1 + l_2) \prod_i \G({\alpha_i \over 2}+1) \G({1 \over 2} \Sigma +{3 \over 2}) \over (l_3+1) \left(  \prod_i^3  \G(l_i -1/2)  \right) |x-y|^{\alpha_3} |y-z|^{\alpha_1} |z-x|^{\alpha_2}} \no
&&\times  ( \delta^{AC} \delta^{BD} - \delta^{AD} \delta^{BC} )C(I_1, I_2,I_3).
\eea
The normalizations of the fields can be fixed by demanding that the two point functions are
\beq
\langle  A^{(I_1)}(x) A^{(I_2)}(y) \rangle =  
{ \delta^{I_1I_2} \over |x-y|^{2l}}.
\eeq
for any two chiral primary fields $A^I$. After rescaling the fields in order to satisfy this condition, the normalized three-point function are given by:
\bea
\langle T^{(1)}(x) T^{(2)}(y) T^{(3)}(z) \rangle &=& -{2^{7/2} \over \pi^{1/4}}
{\prod_i \G({\alpha_i \over 2}+1) \G({1 \over 2} \Sigma 
+{3 \over 2}) \over 
\sqrt{ \prod_i^3 \left( \G(l_i +3/2) \G(l_i+2) (l_i +1) 
(l_i^2-1) \right) } }\no
&&\times\frac{(l_1(l_1^2-1)+l_2 (l_2^2-1) + l_3 (l_3^2-1))}{|x-y|^{\alpha_3} |y-z|^{\alpha_1} |z-x|^{\alpha_2} }C(I_1, I_2,I_3) \no
\no
\langle S^{(1)} S^{(2)} T^{(3)}\rangle   &=&-{2^{5/2} \over \pi^{1/4}}
{\prod_i \G({\alpha_i \over 2}+1 ) \G({1 \over 2} \Sigma +{3 \over 2})
\over 
\sqrt{ \prod_i^3 \left( \G(l_i +3/2) \G(l_i +1) (l_i^2-1)\right)(l_3+1)} }\no
&&\times \frac{(l_1(l_1^2-1)+l_2 (l_2^2-1) + l_3 (l_3^2-1))}{|x-y|^{\alpha_3} |y-z|^{\alpha_1} |z-x|^{\alpha_2}   }   C(I_1, I_2,I_3) \no 
\no
\langle  V^{(1)AB} U^{(2)CD} S^{(3)} \rangle &=&{2^2 \over \pi^{1/4}}
{\prod_i \G({\alpha_i \over 2}+1) \G({1 \over 2} \Sigma +{3 \over 2})(l_1-l_2)  C(I_1, I_2,I_3)(\delta^{AC}\delta^{BD} - \delta^{AD}\delta^{BC})
\over 
\sqrt{\prod_i^3 \left( \G(l_i +3/2)\G(l_i +1) \right)(l_3^2-1)} |x-y|^{\alpha_3} |y-z|^{\alpha_1} |z-x|^{\alpha_2} } \no 
\no 
\langle V^{(1)}_{AB} V^{(2)}_{CD} V^{(3)}_{EF} \rangle   & = &-{2^2  \over \pi^{1/4}}
{\prod_i \G({\alpha_i \over 2}+1) \G({1 \over 2} \Sigma +{3 \over 2}) C(I_1, I_2,I_3)  \epsilon_{12ABCDEF}
\over 
\sqrt{\prod_i^3 \left( \G(l_i +3/2) \G(l_i +1) \right)} |x-y|^{\alpha_3} |y-z|^{\alpha_1} |z-x|^{\alpha_2} }  \no 
\no
\langle V^{(1)}_{AB} V^{(2)}_{CD} T^{(3)} \rangle   &=& -{2^{7/2} \over \pi^{1/4}}
{(l_1 + l_2)\prod_i \G({\alpha_i \over 2}+1) \G({1 \over 2} \Sigma +{3 \over 2}) 
\over 
(l_3+1) \sqrt{ \prod_i^3 \left( \G(l_i +3/2) \G(l_i +1) \right) (l_3^2-1)}  } \no
&& \times \frac{(\delta_{AC} \delta_{BD} - \delta_{AD} \delta_{BC} ) C(I_1, I_2,I_3)   }{|x-y|^{\alpha_3} |y-z|^{\alpha_1} |z-x|^{\alpha_2}  }
  \no 
\no
\langle  U^{(1)}_{AB} U^{(2)}_{CD} V^{(3)}_{EF} \rangle   &=&{2^2 \over \pi^{1/4}}
{ \prod_i \G({\alpha_i \over 2}+1) \G({1 \over 2} \Sigma +{3 \over 2}) C(I_1, I_2,I_3)\epsilon_{12ABCDEF}
\over 
\sqrt{ \left(  \prod_i^3 \G(l_i +3/2) \G(l_i +1)  \right)} |x-y|^{\alpha_3} |y-z|^{\alpha_1} |z-x|^{\alpha_2} } \no 
\no
\langle  U^{(1)AB} U^{(2)CD} T^{(3)} \rangle    &=& -{2^{7/2} \over \pi^{1/4}}
{ (l_1 + l_2) \prod_i \G({\alpha_i \over 2}+1) \G({1 \over 2} \Sigma +{3 \over 2}) 
\over 
(l_3+1) \sqrt{ \left(  \prod_i^3  \G(l_i +3/2) \G(l_i +1) \right) (l_3^2-1)}  }\no
&&\times \frac{( \delta^{AC} \delta^{BD} - \delta^{AD} \delta^{BC} )C(I_1, I_2,I_3)  }{|x-y|^{\alpha_3} |y-z|^{\alpha_1} |z-x|^{\alpha_2}  } .
\eea
\section{Conclusions}

In this paper, we gave the calculation of three point interactions of
chiral primaries in 4D supergravity. In order to remove the derivatives from the interactions, we used 
nontrivial redefinitions of the fields which solved the linear equation
of motion for chiral primaries. It is after removing the derivative 
terms that we deal with a standard field theory in two dimensions. These redefinitions were also needed for the cases of higher dimensional $AdS$ spaces\cite{leeseiberg},\cite{mihail}. Deeper reasons behind these redefinitions are  
still not clear.   \footnote{See also  ref.\cite{nastase} for related discussions for the case of the $D=11$ supergravity on $AdS_7 \times S^4$.} 
 
\par
In section 4, we derived the three point interactions. The factors appear which are similar to the results in higher dimensional $AdS$ spacetimes. In order to make a useful statement on $AdS_2/CFT_1$ correspondence, similar computation has to be made in the boundary conformal quantum mechanics dual to this theory. There has been a conjecture that this dual quantum mechanics is given by a supersymmetric extension of the Calogero model\cite{gibtown},\cite{town},\cite{jy}.  It would be interesting to see whether this is true. However, since a many-body quantum mechanics becomes a $1+1$ dimensional field theory after the second quantization, one might directly obtain this field theory starting from the Lagrangian (\ref{trla}) by appropriate truncation. This interesting issue, along with the evluation of the interaction terms for the fermions, are left for future studies.

\bigskip

\noindent{\bf Acknowledgements:}
 I give special thanks to Antal Jevicki for many helpful discussions and suggestions. I also thank Tamiaki Yoneya and Sangmin Lee for useful discussions. This work was supported by JSPS through Institute of Physics, University of Tokyo.
 
 \bigskip
 \bigskip

\appendix{\bf Appendix 1 : Vector and scalar part of the lagrangian}
As mentioned in the text, the $SU(8)$ gauge symmetry can be fixed to the so called symmetric gauge where 
${\cal V} = \exp(X)$ and 
\beq
X = \pmatrix{
 0 & W_{ABCD}  \cr
 \bar W^{ABCD}  &
 0 }
\eeq
Expanding in $W$, we get the expressions
\beq
{\cal V} = \pmatrix{  {\delta_{AB}}^{CD} + \frac{1}{2}W_{ABEF} \bar W^{EFCD}+O(W^4)  &
 W_{ABCD}+O(W^3) \cr
\bar W^{ABCD} + O(W^3)  &
 {\delta^{AB}}_{CD}+\frac{1}{2}\bar W^{ABEF} W_{EFCD} + O(W^4) 
} ,
\label{s1}
\eeq

and 

\beq 
{\cal V}^{-1} = \pmatrix{
 {\delta_{AB}}^{CD} + \frac{1}{2}W_{ABEF} \bar W^{EFCD}+O(W^4) &
 -W_{ABCD}+O(W^3) \cr
 -\bar W^{ABCD} + O(W^3)   &
 {\delta^{AB}}_{CD}+\frac{1}{2}\bar W^{ABEF} W_{EFCD} + O(W^4)
}. 
\label{s2}
\eeq 

From these expressions one easily gets

\begin{equation}
\partial_{\muhat}{\cal V}{\cal V}^{-1} = \pmatrix{{1 \over 2}\na_\mu (W \bar W)-(\na_\mu W) \bar W +O(W^4) &
 \na_\mu W+O(W^3) \cr
 \na_\mu \bar W + O(W^3)   &
 {1 \over 2}\na_\mu (\bar W W)- (\na_\mu \bar W) W +O(W^4) } 
\label{s3}
\end{equation}

Comparing with Eq.(\ref{Vconnection}), we see that
\beq
P_{\muhat ABCD}  = \na_{\hat \mu} W_{ABCD} + O(W^3), \quad \bar P^{\muhat ABCD}  = \na_{\hat \mu} W^{ABCD} + O(W^3) .
\eeq

$\tilde H(B, {\cal V})$ is defined by the equation

\begin{equation}
\pmatrix{ G_{\muhat \nuhat} + i H_{\muhat \nuhat}  \cr
 G_{\muhat \nuhat} - i H_{\muhat \nuhat}  
}
=  ({\cal V}^\dagger {\cal V})^{-1} \pmatrix{i \t G_{\muhat \nuhat} - \t H_{\muhat \nuhat }   \cr
 -i \t G_{\muhat \nuhat} - \t H_{\muhat \nuhat } },
\label{E7F}
\end{equation}
where it is to be understood that the matrices are multiplied by contracting the $SU(8)$ indices, which I did not write explicitly. \footnote{In this paper, I use only $SU(8)$ indices, in contrast to ref.\cite{corley} where $E_7$ indices were also used.} $\t G, \t H$ denotes the dual fields,
\beq
\t G^{\hat \mu \hat \nu} =  \frac{1}{2} \epsilon^{\hat \mu \hat \nu \hat \rho \hat \sigma} G_{\hat \rho \hat \sigma}, \quad \t H^{\hat \mu \hat \nu} =  \frac{1}{2} \epsilon^{\hat \mu \hat \nu \hat \rho \hat \sigma} H_{\hat \rho \hat \sigma}
\eeq
Expanding in $W$, we get
\begin{equation}
\pmatrix{G_{\muhat \nuhat} + i H_{\muhat \nuhat}  \cr
 G_{\muhat \nuhat} - i H_{\muhat \nuhat}}
=  \pmatrix{1 + 2 W \bar W +O(W^4)  &
 -2 W -\frac{4}{3} W \bar W W + O(W^5)  \cr
-2 \bar W -\frac{4}{3} \bar W W \bar W + O(W^5)   &
1 + 2 \bar W W +O(W^4) }
 \pmatrix{i \t G_{\muhat \nuhat} - \t H_{\muhat \nuhat }   \cr
-i \t G_{\muhat \nuhat} - \t H_{\muhat \nuhat } } .
\label{v2}
\end{equation}
By adding the first and second row, we get
\bea
G&=&-(1-W-\bar W + W \bar W + \bar W W -{2 \over 3} \bar W W \bar W -{2 \over 3} W \bar W W + O(W^5) ) \t H \no
 &&+ i (- \bar W + W + W \bar W -  \bar W W -{2 \over 3} \bar W W \bar W +{2 \over 3} W \bar W W ) \t G.
\eea 
Solving in terms of $H$, we finally get
\bea
G^{AB}_{\muhat \nuhat} \tilde{H}^{(B)}_{\muhat \nuhat AB} &=& 
-G^{AB}_{\muhat \nuhat} (1+W+\bar{W}+W^2+\bar{W}^2 \no 
&&-{1 \over 3} W \bar{W} W -{1 \over 3} \bar{W} W \bar{W} +  W^3 + \bar {W}^3 )_{ABCD}  G^{\muhat \nuhat CD} \no 
&&+ i G^{AB}_{\muhat \nuhat}(W-\bar{W}+W^2-\bar{W}^2+ O(W^3) )_{ABCD} 
\tilde{G}^{\muhat \nuhat CD} + O(W^4) , \no
. 
\eea

\appendix{\bf Appendix 2 : Spherical Harmonics}

 We list some formulas about spherical harmonics needed for the calculations in the text. The spherical hamonics on $S^2$ are very simple in that there are only scalar spherical harmonics, which we denote by $Y_I = Y_{lm}$. They are normalized so that
\beq
\int Y_{l_1 m_1} Y_{l_2m_2} = \delta_{l_1 l_2} \delta_{m_1 m_2}.
\eeq
The explicit form is given by
\beq
Y_{l m} \equiv {1 \over 2} \sqrt{(2l+1)(l-m) \over\pi (l+m)}e^{i m \phi} P_l^m (cos \theta)
\eeq
where $P_l^m(x)$ is the associated Legendre polynomial
\beq
P_l^m(x) \equiv {(-1)^m \over 2^l l!} (1-x^2)^{m/2} {d^{m+l} \over dx^{m+l}}(x^2-1)^l,
\eeq
although this explicit form was not really used in the text. More important is the formula for the integral of three spherical harmonics with derivatives, expressed in terms of $C(I_1, I_2, I_3)$. They are as follows: \footnote{again, we use the abbreviation $i$ for   $I_i \equiv (l_i, m_i)$}
\begin{mathletters}
\bea
A(1;2,3) &\equiv& \int Y_1 \na Y_2 \na Y_3 \no
&=&{1 \over 2}(l_2(l_2+1) + l_3 (l_3+1) - l_1 ( l_1+1))C(1,2,3) \label{arule} \\
B(1;2,3) &\equiv& \int Y_1 \na_\alpha \na_\beta Y_2  \na^\alpha \na^\beta Y_3 \no
&=& {1 \over 4} (l_1(l_1+1) - l_2(l_2+1)-l_3(l_3+1))\no
&&\times (2+l_1(l_1+1) - l_2(l_2+1)-l_3(l_3+1)) C(1,2,3) \label{brule} \\
D(1;2,3) &\equiv& \int \na_\alpha \na_\beta Y_1 \na^\alpha  Y_2  \na^\beta Y_3 \no
&=& {1 \over 4} (l_3(l_3+1)+l_1(l_1+1) - l_2(l_2+1))\no
&&\times (l_3(l_3+1)+l_2 (l_2+1) - l_1(l_1+1)) C(1,2,3) \label{drule} \\
G(1,2,3) &\equiv& \int \na_\alpha \na_\beta Y_1 \na_\beta \na_\gamma Y_2  \na^\gamma \na^\alpha Y_3 \no
&=& {1 \over 8}     ( - l_1^3(l_1+1)^3- l_2^3(l_2+1)^3- l_3^3(l_3+1)^3 \no
&& -2 l_1^2 (1_1+1)^2  -2 l_2^2 (1_2+1)^2   -2 l_3^2 (1_3+1)^2  \no
&&  + 4 l_1 l_2(l_1+1)(l_2+1)+ 4 l_2 l_3(l_2+1)(l_3+1)+ 4 l_3 l_1(l_3+1)(l_1+1) \no
&&+l_1^2 l_2(l_1+1)^2(l_2+1)  +l_1 l_2^2(l_1+1)(l_2+1)^2  +l_2^2 l_3(l_2+1)^2(l_3+1)  \no
&&+l_2 l_3^2(l_2+1)(l_3+1)^2  +l_3^2 l_1(l_3+1)^2(l_1+1)  +l_3 l_1^2(l_3+1)(l_1+1)^2    \no
&& - 2 l_1 l_2l_3  (l_1+1)( l_2 +1)(l_3+1)       ) \label{grule}
\eea 
\end{mathletters}

\noindent
proof)
 
\noindent
a) By integrating by parts, one gets
\bea
A(1;2,3)&=& -\int Y_1 \na^2 Y_2 Y_3 - \int \na_\alpha Y_1 \na^\alpha Y_2  Y_3 \no
&=& l_2 (l_2 +1) C(1,2,3) -A(3;1,2).
\eea 
By permuting the indices and adding the resulting equations, we get Eq.(\ref{arule}).

\noindent
b) By integrating by parts, we get
\bea
B(1;2,3) &=& \int Y_1 \na_\alpha \na_\beta Y_2  \na^\alpha \na^\beta Y_3 \no
&=& -\int \na_\alpha Y_1 \na^\alpha \na^\beta Y_2 \na_\beta Y_3  - \int Y_1 \na_\alpha \na_\beta \na^\alpha Y_2  \na^\beta Y_3 \no
&=&  \int \na_\alpha Y_1 \na^\alpha Y_2 \na_y^2 Y_3 + \na_\beta \na_\alpha Y_1 \na^\alpha Y_2 \na^\beta Y_3 \no
&& -Y_1 \na^\gamma Y_2 \na^\beta Y_3 R_{\beta \gamma} - Y_1 \na^\beta \na_y^2 Y_2 \na_\beta Y_3 \no
&=& -l_3 (l_3 +1) \na_\alpha Y_1 \na^\alpha Y_2 Y_3 - \na_\beta \na_\alpha \na^\beta Y_1 \na^\alpha Y_2 Y_3 - \na_\beta \na_\alpha Y_1\na^\beta \na^\alpha Y_2 Y_3 \no
&& - Y_1 \na^\beta Y_2 \na_\beta Y_3 + l_2 (l_2+1) Y_1 \na^\beta Y_2 \na_\beta Y_3 \no
&=&  -l_3 (l_3+1) A(3;1,2) -  R_{\alpha \gamma} \na^\alpha Y_1 \na^\gamma Y_2 Y_3 \no
&& - \na_\alpha \na_y^2 Y_1 \na^\alpha Y_2 Y_3 - B(3; 1,2) \no
&& + A(1;2,3)(l_2(l_2+1)-1) \no
&=& (l_1(l_1+1)-l_3 (l_3+1)-1) A(3;1,2)  - B(3; 1,2) + A(1;2,3)(l_2(l_2+1)-1)
\eea
which gives 
\beq
B(1;2,3)+B(3; 1,2)=(l_2(l_2+1)-1)  A(1;2,3)+(l_1(l_1+1)-l_3 (l_3+1)-1) A(3;1,2)\label{b1} 
\eeq
On the other hand, we have
\bea
B(1;2,3) &=& -\int \na^\alpha Y_1 \na_\alpha \na_\beta Y_2  \na^\beta Y_3 -\int  Y_1 \na^\alpha \na_\beta \na_\alpha Y_2  \na^\beta Y_3 \no
&=& \int \na^\alpha \na^\beta Y_1 \na_\alpha \na_\beta Y_2  Y_3 +\int  \na^\alpha Y_1 \na^\beta \na_\alpha \na_\beta  Y_2  Y_3 \no
&&- R_{\beta \gamma}  \int Y_1 \na^\gamma Y_2 \na^\beta Y_3 -  Y_1 \na^\beta \na_y^2 Y_2 \na_\beta Y_3 \no
&=& B(3;1,2)+ R_{\alpha \gamma} \na^\alpha Y_1 \na^\gamma Y_2 Y_3 \no
&& + \na_\alpha Y_1 \na^\alpha \na_y^2 Y_2 Y_3 - Y_1 \na^\beta Y_2 \na_\beta Y_3 + l_2 (l_2+1) Y_1 \na_\beta Y_2 \na^\beta Y_3 \no
&=& B(3;1,2)+ \na^\alpha Y_1 \na_\alpha Y_2 Y_3 \no
&& -l_2 (l_2+1) \na_\alpha Y_1 \na^\alpha  Y_2 Y_3 - Y_1 \na^\beta Y_2 \na_\beta Y_3 + l_2 (l_2+1) Y_1 \na_\beta Y_2 \na^\beta Y_3 \no
&=&B(3;1,2) + (1-l_2(l_2+1))(A(3;1,2)  -  A(1;2,3))
\eea
which is
\beq
B(1;2,3)-B(3;1,2) = (1-l_2(l_2+1))(A(3;1,2)  -  A(1;2,3)) \label{b2}
\eeq
Adding (\ref{b1}) and (\ref{b2}) gives Eq.(\ref{brule}).

\noindent
c) 
\bea
D(1;2,3) &=& \int \na_\alpha \na_\beta Y_1 \na^\alpha  Y_2  \na^\beta Y_3 \no
&=& - \int \na_\beta Y_1 \na_y^2 Y_2 \na^\beta Y_3 - \na^\beta Y_1 \na^\alpha Y_2 \na_\alpha \na_\beta Y_3 \no
&=& l_2(l_2+1) \int \na_\beta Y_1 Y_2 \na^\beta Y_3 + Y_1 \na_\beta \na_\alpha Y_2 \na^\beta \na^\alpha Y_3 + Y_1 \na_\alpha Y_2 \na^\beta \na^\alpha \na_\beta Y_3 \no
&=& l_2(l_2+1) \int \na_\beta Y_1 Y_2 \na^\beta Y_3 + Y_1 \na_\beta \na_\alpha Y_2 \na^\beta \na^\alpha Y_3 \no
&&+ R_{\alpha \gamma} Y_1 \na^\alpha Y_2 \na^\gamma Y_3   + Y_1 \na_\alpha Y_2  \na^\alpha \na_y^2 Y_3 \no
&=& l_2 (l_2+1) A(2;3,1) + B(1; 2,3) + (1-l_3(l_3+1))A(1;2,3) 
\eea
which is equivalent to Eq.(\ref{drule}).

\noindent
d)
\bea
G(1,2,3) &=& \int \na_\alpha \na_\beta Y_1 \na_\beta \na_\gamma Y_2  \na^\gamma \na^\alpha Y_3 \no
&=& - \int \na_\gamma \na_\alpha Y_1 \na^\alpha \na^\beta \na^\gamma Y_2 \na_\beta Y_3 - (1-l_1 (l_1+1))\int \na_\gamma  Y_1 \na^\beta \na^\gamma Y_2 \na_\beta Y_3 \no
&=& - \int \na_\gamma \na_\alpha Y_1({{{R_\alpha}^\beta}_\gamma}^\delta  \na_\delta Y_2 + \na^\beta \na_\alpha \na^\gamma Y_2) \na_\beta Y_3    +  (l_1 (l_1+1)-1)\int \na_\gamma  Y_1 \na^\beta \na^\gamma Y_2 \na_\beta Y_3 \no
&=& (l_1(l_1+1) -1)D(2;3,1) - \int \na_y ^2 Y_1 \na_\beta Y_2 \na^\beta Y_3 \no &&+ \int \na^\beta \na^\alpha Y_1 \na_\alpha Y_2 \na_\beta Y_3 - \int \na_\gamma \na_\alpha Y_1 \na^\beta \na_\alpha \na^\gamma Y_2 \na_\beta Y_3 \no
&=&  (l_1(l_1+1) -1)D(2;3,1) +l_1(l_1+1) A(1;2,3) + D(1; 2,3)-l_3 (l_3 +1) B(3;1,2) \no 
&&+ \int \na^\beta \na_\gamma \na_\alpha Y_1 \na^\alpha \na^\gamma  Y_2 \na_\beta Y_3 \no
&=&   (l_1(l_1+1) -1)D(2;3,1) +l_1(l_1+1) A(1;2,3) + D(1; 2,3)-l_3 (l_3 +1) B(3;1,2) \no 
&&+ \int ( {{R^\beta}_{\alpha \gamma}}^\delta \na_\delta Y_1+\na_\alpha \na^\beta \na_\gamma Y_1 )\na^\alpha \na^\gamma  Y_2 \na_\beta Y_3 \no
&=&   l_1(l_1+1) D(2;3,1) +l_1(l_1+1) A(1;2,3) -l_3 (l_3 +1) B(3;1,2) \no 
&&  + l_2 (l_2 +1) A(2; 3,1) + l_2 (l_2 +1) D(1; 2,3)
\eea
which is Eq.(\ref{grule}).

\end{document}